\documentclass[aps,twocolumn,amsmath,amssymb,preprintnumbers]{revtex4}
\usepackage{amsmath} \usepackage{amsfonts} \usepackage{amssymb}
\usepackage{titlesec}
\usepackage{bbm}
\usepackage{epsfig}
\usepackage{graphics}
\usepackage{graphicx}
\newcommand{\be}{\begin{equation}}
\newcommand{\ee}{\end{equation}}
\newcommand{\ba}{\begin{eqnarray}}
\newcommand{\ea}{\end{eqnarray}}
\newcommand{\nn}{\nonumber}
\newcommand{\kl}{\langle}
\newcommand{\kr}{\rangle}

\newcommand{\G}{\Gamma^{(2)}}
\newcommand{\vk}{\vec k}
\newcommand{\w}{w^\pm_k}
\newcommand{\ww}{w^+_k}
\newcommand{\www}{w^-_k}
\newcommand{\et}{(\eta)}
\newcommand{\kk}{(k)}

\titleformat{\subsection}[block]{\normalfont\bfseries}{\thesubsection.}{1ex}{}
\titlespacing{\subsection}{0pt}{10pt}{1pt}[0pt]
\titleformat*{\section}{\large\bfseries}
\renewcommand{\thesubsection}{\arabic{subsection}}

\begin{document}

\title[ ]{Can observations look back to the beginning of inflation?}

\author{C. Wetterich}
\affiliation{Institut  f\"ur Theoretische Physik\\
Universit\"at Heidelberg\\
Philosophenweg 16, D-69120 Heidelberg}

\begin{abstract}
The cosmic microwave background can measure the inflaton potential only if inflation lasts sufficiently long before the time of horizon crossing of observable fluctuations, such that non-linear effects in the time evolution of Green's functions lead to a loss of memory of initial conditions for the ultraviolet tail of the spectrum. Within a derivative expansion of the quantum effective action for an interacting scalar field we discuss the most general solution for the correlation function, including arbitrary pure and mixed quantum states. In this approximation no loss of memory occurs - cosmic microwave observations see the initial spectrum at the beginning of inflation, processed only mildly by the scale-violating effects at horizon crossing induced by the inflaton potential. 
\end{abstract}

\maketitle

Has the universe at the beginning of inflation been in a particular vacuum state, or can fluctuations have started with a more general, perhaps even ``classical'' state? Can we detect the difference by observation? The usual treatment of primordial fluctuations \cite{MUK,STAT,RSV,STA,GP,BST,AW} assumes that fluctuations are described by a particular state, the Bunch-Davies vacuum \cite{BD}. Investigations of more general initial conditions \cite{AEH}, based on pioneering work on the time evolution of correlation functions for free quantum fields \cite{VF}, seem to suggest that a particular universal state is approached independently of initial conditions. (For early work see, e.g. refs. \cite{TH1,TH2,BH}.) Similar observations \cite{AHR} for coarse grained bulk quantities as the energy momentum tensor apparently point to a similar direction. In contrast, many authors implicitly assume that the primordial cosmic fluctuations keep memory of initial conditions, see e.g. ref. \cite{BM}.

We investigate here the effect of initial information on the observable fluctuation spectrum. For this purpose we employ an evolution equation for the correlation function which is exact for a given form of the quantum effective action. The operator formalism and the explicit choice of a vacuum are not needed. We discuss interacting scalar fields in an arbitrary homogeneous and isotropic background geometry. For an expansion of the effective action up to second order in derivatives we find the general solution for the correlation function, which includes mixed quantum states beyond the usually considered pure states. In this approximation the memory of the initial spectrum remains imprinted on the observable spectrum. 

It is instructive to divide the time evolution of the correlation function into two epochs. For the first epoch the wavelength of the relevant fluctuations remains well inside the horizon. The second epoch covers the period of horizon crossing and afterwards. Only for the second period the inflaton potential will play a role. The fluctuation spectrum at the end of the first period is called ``pre-spectrum''. More precisely, we define the ``pre-spectrum'' as the equal-time correlation function a couple of $e$-foldings before horizon crossing. The detailed geometry and inflaton potential in the epoch of horizon crossing only process the pre-spectrum, modifying it by small scale-symmetry violating effects. 

If the pre-spectrum keeps memory of the fluctuations in the ``initial state'' of the universe at the beginning of inflation, the observation of amplitude and spectral index of the fluctuation spectrum, as imprinted on the cosmic microwave background, yields information on the beginning of inflation. The relation between fluctuation amplitude $\Delta^2(k)$ and inflaton potential $V$ at horizon crossing is modified by a factor $(A_p+1),A_p\geq 0$, as compared to the usual result of the slow roll approximation which corresponds to $A_p=0$, i.e.
\be\label{1}
\Delta^2=\frac{(A_p+1)V}{24\pi^2\epsilon M^4}.
\ee
The enhancement $A_p+1$ could be large. Also the spectral index $n_s$ receives an additional contribution $n_p$ from the pre-spectrum
\be\label{2}
n_s=1+n_p-6\epsilon+2\eta.
\ee
Here $M$ is the reduced Planck mass and $\epsilon, \eta$ are the slow roll parameters. 

In the absence of interactions the pre-spectrum keeps the full memory of the initial spectrum at the beginning of inflation. We include the effects of interactions within a formalism based on the quantum effective action $\Gamma$. Its second functional derivative $\Gamma^{(2)}$ yields the exact inverse propagator, such that the correlation function (connected two point function, Green's functions, propagator) $G$ has to obey
\be\label{3}
\Gamma^{(2)}G=1.
\ee
Since $\Gamma^{(2)}$ contains time derivatives, this defines an exact time-evolution equation for the correlation function. For a known or assumed form of $\G$ it can be followed from an initial state of the fluctuations at the beginning of inflation until horizon crossing. 

We consider a scalar inflaton field in an arbitrary homogeneous and isotropic geometry, characterized by a scale factor $a(\eta)$ with conformal time $\eta$. We also account for an arbitrary evolution of the homogeneous inflaton mean field $\bar\varphi(\eta)$. We find that no loss of memory occurs as long as the following three approximations on the form of the effective action remain valid: (i) The effective action $\Gamma$ can be characterized for all times after the beginning of inflation by its universal ``no boundary form'' where it has no explicit dependence on position and time. General covariance and locality (derivative expansion) restrict then its possible form to a few relevant invariants. (ii) Backreaction effects do not induce substantial non-linearities via the impact of $G$ on the time evolution of the background fields. (iii) For technical simplicity we first assume the validity of a derivative expansion of the form
\be\label{4}
\Gamma=\int_x\sqrt{g}
\left\{\frac12 \partial^\mu\varphi \partial_\mu\varphi+{\cal U}(\varphi;\eta)\right\}, 
\ee
with $\sqrt{g}=i\sqrt{-\det (g_{\mu\nu})}$. The non-linearities due to interactions are accounted for by the effective potential ${\cal U}$.
The dependence of ${\cal U}$ on background fields can make the effective action for the fluctuations explicitly time-dependent. 

Under these general assumptions we can ``see'' extreme sub-Planckian information \cite{BM} at the beginning of inflation! The memory of initial conditions is not lost and the pre-spectrum reflects the initial spectrum at the beginning of inflation.

Possible effects for observation may be demonstrated by one of the solutions of the evolution equation for the correlation function that corresponds to a mixed quantum state. For this particular solution we consider an initial spectrum that differs from the ``Bunch-Davies spectrum'' by a factor $1+A_p(k)$. We will show below that the same factor appears in the pre-spectrum. We briefly discuss two somewhat arbitrary but instructive examples. 

For our first example we take
\be\label{5}
A_p(k)=\frac{A}{2}\left(1-\frac2\pi arctg~ x\right)~,~x=B^{-1}\ln \left(\frac{k}{k_0}\right),
\ee
with $k$ the comoving wave number. In the ultraviolet (UV) limit $k\to \infty$ the modification of the Bunch-Davies spectrum vanishes, while for $k\ll k_0$ the multiplicative factor is a constant $A$. Values $A\gg 1$ may be associated with ``classical fluctuations''.

The vanishing of $A_p(k)$ for $k\to \infty$ is motivated by the consideration that some type of equilibration effect not contained in our approximation may drive the propagator for the large momentum fluctuations towards the Lorentz-invariant vacuum propagator in flat space.

Any modification of a pre-spectrum with $A_p(k)\neq 0$ and $A_p(k\to\infty)=0$ will involve a violation of scale symmetry. This entails the appearance of some fixed scale $k_0$ where in our case the UV-behavior switches to the IR-behavior. Finally, the parameter $B$ describes the width of the UV-IR-crossover in logarithmic $k$-space. The contribution of such a pre-spectrum to the spectral index $n_s$ reads
\be\label{6}
n_p=-\left[B(1+x^2)\left(\frac{\pi}{2}-{arctg}~ x+\frac\pi A\right)\right]^{-1}.
\ee
One has 
\be\label{EA}
n_p\big(x=0(1)\big)=-\frac{2}{\pi B} \left(1+\frac{2(4)}{A}\right)^{-1},
\ee
while for $|x|\gg1$ the modification $n_p$ becomes small, suppressed by powers of $|x|$. For large $A$ the corrections $\sim A^{-1}$ can be neglected and $n_p$ is suppressed only by the width $B^{-1}$ if $x^2\lesssim 2$. A small value $A\ll 1 $ appears as an additional suppression factor. 

If we associate at the beginning of inflation the scale of UV-IR-crossover with the Planck mass, the scale $k_0$ is determined such that 
\be\label{8}
x=B^{-1}\left(N_{in}-\ln \left(\frac{M}{H_0}\right)\right),
\ee
with $N_{in}$ the number of $e$-foldings between the beginning of inflation and horizon crossing. For a narrow crossover with $B$ of the order one a substantial  modification of the observable spectral index would occur only if $N_{in}$ is in the vicinity of $\ln (M/H_0)$. In contrast, for a wide crossover $B\gg 1$ a much wider range in $N_{in}$ would lead to an observable trace in the spectrum. Except for special situations we conclude from eq. \eqref{6} that the modification of the spectral index remains small. The memory of the initial spectrum will not change the observed approximate scale invariance of the primordial spectrum. Only the precise relation between the observed spectrum and the inflaton potential is at stake. The amplitude is of the order $A+1$ unless $N_{in}/B\gg 1$ and can experience a substantial enhancement. In this case the value of the inflaton potential at horizon crossing would be much smaller than inferred usually from the normalization of the CMB anisotropies. 

We emphasize that the observable modes have all been inside the horizon at the beginning of inflation. The larger $N_{in}$, the more the ultraviolet tail of the initial spectrum is explored. Our example is thought to cover only the range of $k$ for modes inside the horizon at the beginning of inflation. For smaller $k$ the initial power spectrum may actually vanish without affecting our discussion.

For our second example we abandon the property $A_p(k\to\infty)\to 0$. We take
\be\label{8A}
1+A_p(k)=A\left(\frac{k}{k_0}\right)^\alpha.
\ee
In this case the contribution of the pre-spectrum to the spectral index becomes simply 
\be\label{8B}
n_p=\alpha.
\ee
It is only small for small $\alpha$. The amplitude proportional to $A$ could be substantially larger than the one for Bunch-Davies initial conditions. Under these circumstances predictions from inflation become only possible if additional information about initial conditions is available. Again, eq. \eqref{8A} may only hold for the observable range of modes, with possible modifications in the far ultraviolet or infrared, for example in order to ensure $A_p(k)\geq 0$.

Let us now turn to the evolution equation for the correlation function and its solution. We will demonstrate that within our three assumptions the memory of initial conditions is not lost. With eq. \eqref{4} and $g_{\mu\nu}=a^2(\eta)\eta_{\mu\nu},\sqrt{g}=ia^4$, the identity \eqref{3} implies for the correlation function in momentum space $G(k,\eta,\eta')$ the exact evolution equation
\ba\label{9}
&&\tilde D_\eta G(k,\eta,\eta')=-\frac{i}{a^2}\delta(\eta-\eta'),\nn\\
&&\tilde D_\eta=\partial_\eta^2+2{\cal H}\partial_\eta+k^2+m^2a^2,
\ea
where
\be\label{10}
{\cal H}(\eta)=\frac{\partial\ln a(\eta)}{\partial\eta}~,~m^2(\eta)=
\frac{\partial^2{\cal U}}{\partial \varphi^2}_{|\bar\varphi(\eta)}.
\ee
The mass term $m^2$ depends on time by virtue of the time-evolving background fields $\bar\varphi$ and possibly other fields. For the particular case of de Sitter space one has $a=-1/(H_0\eta)~,~{\cal H}=-1/\eta$, with $-\infty<\eta<0$. The conformal time $\eta$ can be associated with physical time as measured by the number of oscillations of wave functions, such that $\eta\to-\infty$ actually corresponds to the infinite past of an eternal universe \cite{CWEU}. 

The propagator equation \eqref{9} is the only input for our discussion. The inhomogeneous term on the r.h.s. replaces the usual treatment of commutator relations for quantum field operators. The general solution to the inhomogeneous differential equation \eqref{9} can be found \cite{CWCF} by writing
\ba\label{11}
&&G(k,\eta,\eta')=G_s(k,\eta,\eta')\\
&&\hspace{0.5cm}+G_a(k,\eta,\eta')\big[\theta(\eta-\eta')-\theta(\eta'-\eta)\big],\nn
\ea
with $G_s(k,\eta',\eta)=G_s(k,\eta,\eta'),G_a(k,\eta',\eta)=-G_a(k,\eta,\eta')$ obeying
\be\label{12}
\tilde D_\eta G_s=0~,~\tilde{D}_{\eta}G_a=0~,~\partial_\eta G_{a_{|\eta=\eta'}}=-\frac{i}{2a^2}.
\ee

We observe that the symmetric part $G_s$ obeys a homogeneous linear equation, such that its amplitude is not fixed. The amplitude of the observable power spectrum is directly related to the equal time correlation function $G(k,\eta)=G_s(k,\eta,\eta)$ at horizon crossing,
\be\label{13}
\Delta^2(k)\approx \frac{k^3H^2}{4\pi^2\dot{\bar \varphi}^2}
G(k,\eta)_{|hc}=A_s
\left(\frac{k}{k_s}\right)^{n_s-1},
\ee
with spectral index $n_s$, pivot scale $k_s$ and amplitude $A_s$. Already at this stage we see that the evolution keeps memory of the overall amplitude of the initial fluctuations. As long as no non-linearities in $G$ are introduced this feature holds in a much wider setting including a more complicated form of the scalar kinetic term and beyond the approximation of only two derivatives \cite{CWCF}. 

One may either construct directly the most general solution \cite{CWCF} of eq. \eqref{12} as a linear combination of products of mode fluctuations $w^\pm_k(\eta)$ that obey $Dw=0$. As an equivalent, perhaps more intuitive alternative one may derive from eq. \eqref{12} the evolution equation for the equal time correlation function. The time evolution for equal time correlation functions can be described by three connected two point functions
\ba\label{14}
\kl \varphi(\eta,\vec k)\varphi^*(\eta, \vk')\kr_c&=&G_{\varphi\varphi}(k,\eta)\delta(k-k'),\nn\\
Re\big (\kl \partial_\eta\varphi(\eta,\vk)\varphi^*(\eta,\vk')\kr_c\big)&=&G_{\pi\varphi}(k,\eta)\delta(k-k'),\nn\\
\kl \partial_\eta\varphi(\eta,\vk)\partial_\eta\varphi^*(\eta,\vk')\kr_c&=&G_{\pi\pi}(k,\eta)\delta(k-k'),
\ea
with $\delta(k-k')=(2\pi)^3\delta^3(\vk-\vk')$ and $G_{\varphi\varphi}(k,\eta)=G(k,\eta)$. From eq. \eqref{12} one finds for the dimensionless quantities
\be\label{15}
\tilde G_{\varphi\varphi}=2a^2kG_{\varphi\varphi}~,~\tilde G_{\pi\varphi}=2a^2G_{\pi\varphi}~,~\tilde G_{\pi\pi}=\frac{2a^2}{k}G_{\pi\pi}
\ee
the evolution equation \cite{CWCF}
\ba\label{16}
\partial_u\tilde G_{\varphi\varphi}&=&-\frac{2\tilde h}{u}\tilde G_{\varphi\varphi}+2\tilde G_{\pi\varphi},\nn\\
\partial_u\tilde G_{\pi\varphi}&=&\tilde G_{\pi\pi}-\left(1+\frac{\hat m^2}{u^2}\right)\tilde G_{\varphi\varphi},\nn\\
\partial_u\tilde G_{\pi\pi}&=&\frac{2\tilde h}{u}\tilde G_{\pi\pi}-2\left(1+\frac{\hat m^2}{u^2}\right)\tilde G_{\pi\varphi},
\ea
where $u=k\eta~,~\tilde h=-{\cal H}\eta~,~\hat m^2=a^2\eta^2m^2.$ For de Sitter space one has
$u=-{k}/{(aH_0)}~,~\tilde h=1~,~\hat m^2={m^2}/{H^2_0}.$ The simplest way to derive eq. \eqref{16} uses the general expression of $G$ in terms of mode functions. This shows that the result \eqref{16} can simply be obtained by inserting $D\varphi=0$ inside the brackets in eq. \eqref{14}.

Within our three assumptions (i), (ii), (iii) eq. \eqref{16} is exact. Interaction and fluctuation effects (``higher loops'') are already incorporated in the effective action $\Gamma$. This underlines the power of the use of $\Gamma$. Its first functional derivative yields the exact field equations for the mean fields. The second functional variation provides for the exact evolution equation \eqref{3} for the correlation function and therefore the power spectrum. The third variation is directly related to the bispectrum, and similar for higher order correlations.

Eq. \eqref{16} can be solved numerically, but the most important features for our discussion can be seen easily analytically. Horizon crossing occurs for $u=-1$. The pre-spectrum may be defined at $u=-10$, and initial conditions are set at much larger negative $u$. For the relation between the initial spectrum and the pre-spectrum one only needs the period $|u|\geq 10$. For realistic models of inflation one has $\hat m^2\ll 1$. As a result, the mass term is completely negligible for the computation of the pre-spectrum and only makes a small scale-violating correction in the ``processing'' of the pre-spectrum in the range $-10\leq u\leq- 1$. This processing is exactly the same as in the standard computation. The modifications due to this processing are mild and the information about the pre-spectrum remains preserved. This rather obvious property for a free field remains unchanged in the presence of interactions. 

Only for the asymptotic behavior for $u\to 0$ the mass term becomes important. For positive $\hat m^2$ we find an attraction to the Bunch-Davies vacuum up to a free separate normalization for each $k$-mode. This is in line with the findings of ref. \cite{AEH}. It is not relevant for the observable spectrum since it only concerns the behavior of the pure scalar correlation function for modes far outside the horizon. 

For the computation of the pre-spectrum also $\tilde h/u$ is a small quantity. Deviations of $\tilde h$ from one are $\sim \dot H/H^2$, and therefore further suppressed. For $u<-10$ we may therefore set $\tilde h=1,\hat m^2=0$. The general solution for eq. \eqref{16} is then given by the most general propagator for a massless scalar in de Sitter space which is consistent with translation and rotation symmetry, 
\ba\label{19}
&&\tilde G_{\varphi\varphi}=\alpha(k)\left(1+\frac{1}{u^2}\right)+\beta(k)
\left[\left(1-\frac{1}{u^2}\right)\cos (2u)\right.\\
&&\left.+\frac2u\sin (2u)\right]
+\gamma(k)\left[\frac2u\cos (2u)-\left(1-\frac{1}{u^2}\right)\sin (2u)\right].\nn
\ea
For each $k$-mode it has three free integration constants $\alpha(k),\beta(k),\gamma(k)$. They encode the information about the initial fluctuation spectrum at the beginning of inflation. The Bunch-Davies vacuum corresponds to $\alpha=1~,~\beta=0~,~\gamma=0$ for all values of $k$. The qualitative character of the solution remains valid for $u\leq-1$.

The explicit form of eq. \eqref{19} makes it manifest that for $u=-1$ the memory of initial conditions is not lost for the wide class of situations covered by our already rather general setting. It therefore matters to constrain consistent and in some sense ``reasonable'' initial conditions. We take here a conservative approach motivated by the setting for a free scalar quantum field. We write the general solution to the evolution equation \eqref{9} for the unequal time Green's function for $\eta>\eta'$ in the form \cite{CWCF}
\ba\label{20}
G_>(k,\eta,\eta')&=&\frac{\alpha(k)+1}{2}\www\et\ww(\eta')\\\nn
&+&\frac{\alpha (k)-1}{2}\ww\et\www(\eta')\\\nn
&+&\zeta(k)\ww\et\ww(\eta')+\zeta^*(k)\www\et\www(\eta').\nn
\ea
Here the two ``mode functions'' $\w\et$ are introduced as a basis for the general solution to the homogeneous equation
\be\label{21}
\tilde D_\eta\psi_k\et=0~,~\psi_k\et=c_+\ww\et+c_-\www\et.
\ee
They are normalized according to 
\ba\label{22}
&&\www\et=\big (\ww\et\big)^*,\\
&&\partial_\eta\big[\www\et\ww(\eta')-\ww\et\www(\eta')\big]_{|\eta=\eta'}=-\frac{i}{a^2\et},\nn
\ea
such that eq. \eqref{12} is obeyed. For de Sitter geometry the mode functions obey 
\be\label{19A}
\www=\left(a\sqrt{2k}\right)^{-1}\left(1-\frac{i}{u}\right)e^{-iu}.
\ee
We recover for each $k$ the three independent integration constants, with $\beta(k)=2 Re\zeta(k)$, $\gamma\kk=-2i{\mathcal Im}\zeta\kk$. 

Even though we deal with a general interacting theory, we can cast the general solution into the form of a mixed state for a free effective quantum field
\be\label{23}
G_>(k,\eta,\eta')=\sum_i p_i\psi^{(i)}_k\et\big(\psi^{(i)}_k(\eta')\big)^*,
\ee
where $0\leq p_i\leq 1,~\sum\limits_i p_i=1$. This holds provided $\alpha,\beta$ and $\gamma$ obey the inequalities
\be\label{24}
\alpha\kk\geq 1 ~,~\beta^2\kk+\gamma^2\kk\leq \alpha^2\kk-1.
\ee
It is these conditions that we will impose on the initial conditions. If the last inequality is saturated, $\beta^2+\gamma^2=\alpha^2-1$, we deal with a pure effective quantum state where only one of the probabilities $p_i$ differs from zero. Previous work was restricted to such pure states in the context of free fields \cite{AEH,VF,AHR,BM}, which correspond to the so-called ``$\alpha$-vacua'' \cite{AL,EM}. A pure state with $\alpha\gg1$ can be visualized as a highly excited coherent pure quantum state with large occupation number. States with large $\alpha$ and $\beta=\gamma=0$ correspond to incoherent classical states with large occupation number. 

As a further constraint we may impose that the short distance limit $k\to\infty$ of the propagator is the Lorentz-invariant free propagator. This requires $\alpha(k\to\infty)=1$. Our example \eqref{5} is consistent with these constraints, with $\alpha\kk=1+A_p\kk,~\beta\kk=\gamma\kk=0$. Scale symmetry or the $SO(1,4)$ symmetry of de Sitter space are realized if $\alpha,\beta$ and $\gamma$ are independent of $k$. We do not impose such symmetries on the initial conditions since our aim is to investigate if more general initial conditions tend to approach correlation functions with such symmetries. This is not the case for the general solution. 

A key element for the conservation of memory of initial conditions, and also for the effective free quantum field description, is the linearity of the evolution equation \eqref{9}. We may question its validity. After all, in flat Minkowski space the linear equation \eqref{16} is not able to describe thermalization. Infinitely many conserved quantities obstruct the approach to thermal equilibrium \cite{BEWE}, \cite{BW}, \cite{ABW1}. Alike, we find in our approximation infinitely many conserved quantities, namely $\tilde G_{\varphi\varphi} \tilde G_{\pi\pi}-\tilde G_{\pi\varphi}^2$ for each $k$. If we want to describe a possible process of symmetrization \cite{PT}, where the initial state tends to a state with de Sitter symmetry, we need to take account of non-linearities as, for example, the backreaction. The same holds for dissipation effects which finally could induce a loss of memory of the initial information and lead to a universal asymptotic propagator. 

A convenient tool to incorporate the omitted non-linear effects is the two-particle-irreducible effective action where $G$ appears itself as a generalized field, with an effective action that is not quadratic in $G$. In flat space this approach has given a successful description of thermalization \cite{BC}. It has been conjectured \cite{CWCF} that non-linear equilibration drives the short-distance tail of the Green's function towards the Lorentz-invariant propagator in flat space. This would be sufficient to obtain effectively a Bunch-Davies vacuum at later time (cf. $k\to\infty$ in our example \eqref{5}, \eqref{8}.) Still, a crucial quantity will be the time it takes to symmetrize or equilibrate. This may be very large even on a logarithmic scale in view of the rather tiny self-interactions and gravitational interactions in realistic models of inflation. Only if $N_{in}$ exceeds this equilibration time a loss of memory of detailed initial conditions becomes effective. For smaller $N_{in}$ equilibration is ineffective and the modifications by the pre-spectrum \eqref{1}, \eqref{2} imply a loss of predictivity for such inflationary models. Only after clarification of the role of non-linearities a quantitative 
answer to the question in the title can be given. 

A free propagator for ultrashort distances can also arise from an ultraviolet fixed point for the fundamental theory \cite{CWCG}, which necessarily implies scale invariance of the ``initial spectrum''.  

For an understanding of the possible role of non-linearities it is instructive to compare with the fate of our universe in the far distant future if dark energy is realized as a cosmological constant. We are then at the beginning of a new inflationary era, and the correlation functions for our present universe set the initial data for this future epoch. Our world is not the vacuum, such that the initial spectrum differs strongly from the one for the Bunch-Davies vacuum. Nevertheless, for a large enough number of $e$-foldings the relevant short distance part of the spectrum will be well approximated by the Bunch-Davies vacuum.  

The wavelength of modes crossing the horizon after the first seven $e$-foldings, $N_{in}\approx 7$, corresponds to the present size of gravitationally bound structures, e.g. clusters. These objects will not participate in the expansion, however, such that their comoving wave number $k$ increases and they remain within the horizon by a factor $\sim 1000$. The gravitational potential of such objects will not change fundamentally for the next $10^{11}$ yr and even much further. However, the volume fraction occupied by such bound objects decreases, and so does their contribution to the correlation function. For $N_{in}\approx 10^{30}$ bound objects will experience substantial dissipation once protons decay and charged particles annihilate, leaving only neutrinos in bound objects and 
unbound photons. At this time, however, they occupy so little volume that their contribution to the correlation function is negligible. 

With bound structures playing an ever diminishing role the correlation functions will be dominated by the vacuum of our world. Even though this vacuum contains mass scales as the Fermi scale or $\Lambda_{QCD}$, it is invariant under Lorentz transformations. It is plausible and generally assumed that the propagator of a scalar field in such a vacuum is the one for flat space. (A proof would need to show that the time evolution of correlation functions in empty flat space leads to the standard vacuum correlation, similar to the approach to thermal equilibrium.)

For the example of our future universe equilibration of the correlation functions occurs by a combination of non-linear gravitational effects that lead to bound structures and an approach to the Lorentz invariant vacuum correlation for empty flat space. Taking this lesson over to inflation we conclude that for large $N_{in}$ the initial conditions for fluctuations that cross the horizon are set effectively by the propagator in Minkowski space. 

Can we look back to the beginning of inflation? Yes - at least for time scales that correspond to possible equilibration times. What do we expect to see? Most likely just the correlations of Minkowski space. In this case the observable correlations are universal and inflation maintains its predictivity.

\bibliography{can_observations_look_back_to_the_beginning_of_inflation}

\begin{thebibliography}{25}
\expandafter\ifx\csname natexlab\endcsname\relax\def\natexlab#1{#1}\fi
\expandafter\ifx\csname bibnamefont\endcsname\relax
  \def\bibnamefont#1{#1}\fi
\expandafter\ifx\csname bibfnamefont\endcsname\relax
  \def\bibfnamefont#1{#1}\fi
\expandafter\ifx\csname citenamefont\endcsname\relax
  \def\citenamefont#1{#1}\fi
\expandafter\ifx\csname url\endcsname\relax
  \def\url#1{\texttt{#1}}\fi
\expandafter\ifx\csname urlprefix\endcsname\relax\def\urlprefix{URL }\fi
\providecommand{\bibinfo}[2]{#2}
\providecommand{\eprint}[2][]{\url{#2}}

\bibitem[{\citenamefont{Mukhanov and Chibisov}(1981)}]{MUK}
\bibinfo{author}{\bibfnamefont{V.~F.} \bibnamefont{Mukhanov}} \bibnamefont{and}
  \bibinfo{author}{\bibfnamefont{G.~V.} \bibnamefont{Chibisov}},
  \bibinfo{journal}{JETP Lett.} \textbf{\bibinfo{volume}{33}},
  \bibinfo{pages}{532} (\bibinfo{year}{1981}).

\bibitem[{\citenamefont{Starobinsky}(1979)}]{STAT}
\bibinfo{author}{\bibfnamefont{A.~A.} \bibnamefont{Starobinsky}},
  \bibinfo{journal}{JETP Lett.} \textbf{\bibinfo{volume}{30}},
  \bibinfo{pages}{682} (\bibinfo{year}{1979}).

\bibitem[{\citenamefont{Rubakov et~al.}(1982)\citenamefont{Rubakov, Sazhin, and
  Veryaskin}}]{RSV}
\bibinfo{author}{\bibfnamefont{V.}~\bibnamefont{Rubakov}},
  \bibinfo{author}{\bibfnamefont{M.}~\bibnamefont{Sazhin}}, \bibnamefont{and}
  \bibinfo{author}{\bibfnamefont{A.}~\bibnamefont{Veryaskin}},
  \bibinfo{journal}{Phys.Lett.} \textbf{\bibinfo{volume}{B115}},
  \bibinfo{pages}{189} (\bibinfo{year}{1982}).

\bibitem[{\citenamefont{Starobinsky}(1982)}]{STA}
\bibinfo{author}{\bibfnamefont{A.~A.} \bibnamefont{Starobinsky}},
  \bibinfo{journal}{Phys.Lett.} \textbf{\bibinfo{volume}{B117}},
  \bibinfo{pages}{175} (\bibinfo{year}{1982}).

\bibitem[{\citenamefont{Guth and Pi}(1982)}]{GP}
\bibinfo{author}{\bibfnamefont{A.~H.} \bibnamefont{Guth}} \bibnamefont{and}
  \bibinfo{author}{\bibfnamefont{S.}~\bibnamefont{Pi}},
  \bibinfo{journal}{Phys.Rev.Lett.} \textbf{\bibinfo{volume}{49}},
  \bibinfo{pages}{1110} (\bibinfo{year}{1982}).

\bibitem[{\citenamefont{Bardeen et~al.}(1983)\citenamefont{Bardeen, Steinhardt,
  and Turner}}]{BST}
\bibinfo{author}{\bibfnamefont{J.~M.} \bibnamefont{Bardeen}},
  \bibinfo{author}{\bibfnamefont{P.~J.} \bibnamefont{Steinhardt}},
  \bibnamefont{and} \bibinfo{author}{\bibfnamefont{M.~S.}
  \bibnamefont{Turner}}, \bibinfo{journal}{Phys.Rev.}
  \textbf{\bibinfo{volume}{D28}}, \bibinfo{pages}{679} (\bibinfo{year}{1983}).

\bibitem[{\citenamefont{Abbott and Wise}(1984)}]{AW}
\bibinfo{author}{\bibfnamefont{L.}~\bibnamefont{Abbott}} \bibnamefont{and}
  \bibinfo{author}{\bibfnamefont{M.~B.} \bibnamefont{Wise}},
  \bibinfo{journal}{Nucl.Phys.} \textbf{\bibinfo{volume}{B244}},
  \bibinfo{pages}{541} (\bibinfo{year}{1984}).

\bibitem[{\citenamefont{Bunch and Davies}(1978)}]{BD}
\bibinfo{author}{\bibfnamefont{T.}~\bibnamefont{Bunch}} \bibnamefont{and}
  \bibinfo{author}{\bibfnamefont{P.}~\bibnamefont{Davies}},
  \bibinfo{journal}{Proc.Roy.Soc.Lond.} \textbf{\bibinfo{volume}{A360}},
  \bibinfo{pages}{117} (\bibinfo{year}{1978}).

\bibitem[{\citenamefont{Anderson et~al.}(2000)\citenamefont{Anderson, Eaker,
  Habib, Molina-Paris, and Mottola}}]{AEH}
\bibinfo{author}{\bibfnamefont{P.~R.} \bibnamefont{Anderson}},
  \bibinfo{author}{\bibfnamefont{W.}~\bibnamefont{Eaker}},
  \bibinfo{author}{\bibfnamefont{S.}~\bibnamefont{Habib}},
  \bibinfo{author}{\bibfnamefont{C.}~\bibnamefont{Molina-Paris}},
  \bibnamefont{and} \bibinfo{author}{\bibfnamefont{E.}~\bibnamefont{Mottola}},
  \bibinfo{journal}{Phys.Rev.} \textbf{\bibinfo{volume}{D62}},
  \bibinfo{pages}{124019} (\bibinfo{year}{2000}), \eprint{gr-qc/0005102}.

\bibitem[{\citenamefont{Vilenkin and Ford}(1982)}]{VF}
\bibinfo{author}{\bibfnamefont{A.}~\bibnamefont{Vilenkin}} \bibnamefont{and}
  \bibinfo{author}{\bibfnamefont{L.}~\bibnamefont{Ford}},
  \bibinfo{journal}{Phys.Rev.} \textbf{\bibinfo{volume}{D26}},
  \bibinfo{pages}{1231} (\bibinfo{year}{1982}).

\bibitem[{\citenamefont{Traschen and Hill}(1985)}]{TH1}
\bibinfo{author}{\bibfnamefont{J.~H.} \bibnamefont{Traschen}} \bibnamefont{and}
  \bibinfo{author}{\bibfnamefont{C.~T.} \bibnamefont{Hill}}
  (\bibinfo{year}{1985}), \eprint{FERMILAB-PUB-85-104}.

\bibitem[{\citenamefont{Traschen and Hill}(1986)}]{TH2}
\bibinfo{author}{\bibfnamefont{J.~H.} \bibnamefont{Traschen}} \bibnamefont{and}
  \bibinfo{author}{\bibfnamefont{C.~T.} \bibnamefont{Hill}},
  \bibinfo{journal}{Phys. Rev.} \textbf{\bibinfo{volume}{D33}},
  \bibinfo{pages}{3519} (\bibinfo{year}{1986}).

\bibitem[{\citenamefont{Brandenberger and Hill}(1986)}]{BH}
\bibinfo{author}{\bibfnamefont{R.~H.} \bibnamefont{Brandenberger}}
  \bibnamefont{and} \bibinfo{author}{\bibfnamefont{C.~T.} \bibnamefont{Hill}},
  \bibinfo{journal}{Phys. Lett.} \textbf{\bibinfo{volume}{B179}},
  \bibinfo{pages}{30} (\bibinfo{year}{1986}).

\bibitem[{\citenamefont{Albrecht et~al.}(2015)\citenamefont{Albrecht, Holman,
  and Richard}}]{AHR}
\bibinfo{author}{\bibfnamefont{A.}~\bibnamefont{Albrecht}},
  \bibinfo{author}{\bibfnamefont{R.}~\bibnamefont{Holman}}, \bibnamefont{and}
  \bibinfo{author}{\bibfnamefont{B.~J.} \bibnamefont{Richard}},
  \bibinfo{journal}{Phys.Rev.} \textbf{\bibinfo{volume}{D91}},
  \bibinfo{pages}{043517} (\bibinfo{year}{2015}), \eprint{1410.2612}.

\bibitem[{\citenamefont{Brandenberger and Martin}(2013)}]{BM}
\bibinfo{author}{\bibfnamefont{R.~H.} \bibnamefont{Brandenberger}}
  \bibnamefont{and} \bibinfo{author}{\bibfnamefont{J.}~\bibnamefont{Martin}},
  \bibinfo{journal}{Class.Quant.Grav.} \textbf{\bibinfo{volume}{30}},
  \bibinfo{pages}{113001} (\bibinfo{year}{2013}), \eprint{1211.6753}.

\bibitem[{\citenamefont{Wetterich}(2014{\natexlab{a}})}]{CWEU}
\bibinfo{author}{\bibfnamefont{C.}~\bibnamefont{Wetterich}},
  \bibinfo{journal}{Phys.Rev.} \textbf{\bibinfo{volume}{D90}},
  \bibinfo{pages}{043520} (\bibinfo{year}{2014}{\natexlab{a}}),
  \eprint{1404.0535}.

\bibitem[{\citenamefont{Wetterich}(2015)}]{CWCF}
\bibinfo{author}{\bibfnamefont{C.}~\bibnamefont{Wetterich}}
  (\bibinfo{year}{2015}), \eprint{1503.07860}.

\bibitem[{\citenamefont{Allen}(1985)}]{AL}
\bibinfo{author}{\bibfnamefont{B.}~\bibnamefont{Allen}},
  \bibinfo{journal}{Phys. Rev.} \textbf{\bibinfo{volume}{D32}},
  \bibinfo{pages}{3136} (\bibinfo{year}{1985}).

\bibitem[{\citenamefont{Mottola}(1985)}]{EM}
\bibinfo{author}{\bibfnamefont{E.}~\bibnamefont{Mottola}},
  \bibinfo{journal}{Phys.Rev.} \textbf{\bibinfo{volume}{D31}},
  \bibinfo{pages}{754} (\bibinfo{year}{1985}).

\bibitem[{\citenamefont{Bettencourt and Wetterich}(1998)}]{BEWE}
\bibinfo{author}{\bibfnamefont{L.~M.} \bibnamefont{Bettencourt}}
  \bibnamefont{and}
  \bibinfo{author}{\bibfnamefont{C.}~\bibnamefont{Wetterich}},
  \bibinfo{journal}{Phys.Lett.} \textbf{\bibinfo{volume}{B430}},
  \bibinfo{pages}{140} (\bibinfo{year}{1998}), \eprint{hep-ph/9712429}.

\bibitem[{\citenamefont{Bonini and Wetterich}(1999)}]{BW}
\bibinfo{author}{\bibfnamefont{G.~F.} \bibnamefont{Bonini}} \bibnamefont{and}
  \bibinfo{author}{\bibfnamefont{C.}~\bibnamefont{Wetterich}},
  \bibinfo{journal}{Phys.Rev.} \textbf{\bibinfo{volume}{D60}},
  \bibinfo{pages}{105026} (\bibinfo{year}{1999}), \eprint{hep-ph/9907533}.

\bibitem[{\citenamefont{Aarts et~al.}(2000)\citenamefont{Aarts, Bonini, and
  Wetterich}}]{ABW1}
\bibinfo{author}{\bibfnamefont{G.}~\bibnamefont{Aarts}},
  \bibinfo{author}{\bibfnamefont{G.~F.} \bibnamefont{Bonini}},
  \bibnamefont{and}
  \bibinfo{author}{\bibfnamefont{C.}~\bibnamefont{Wetterich}},
  \bibinfo{journal}{Nucl.Phys.} \textbf{\bibinfo{volume}{B587}},
  \bibinfo{pages}{403} (\bibinfo{year}{2000}), \eprint{hep-ph/0003262}.

\bibitem[{\citenamefont{Berges et~al.}(2005)\citenamefont{Berges, Borsanyi, and
  Wetterich}}]{PT}
\bibinfo{author}{\bibfnamefont{J.}~\bibnamefont{Berges}},
  \bibinfo{author}{\bibfnamefont{S.}~\bibnamefont{Borsanyi}}, \bibnamefont{and}
  \bibinfo{author}{\bibfnamefont{C.}~\bibnamefont{Wetterich}},
  \bibinfo{journal}{Nucl.Phys.} \textbf{\bibinfo{volume}{B727}},
  \bibinfo{pages}{244} (\bibinfo{year}{2005}), \eprint{hep-ph/0505182}.

\bibitem[{\citenamefont{Berges and Cox}(2001)}]{BC}
\bibinfo{author}{\bibfnamefont{J.}~\bibnamefont{Berges}} \bibnamefont{and}
  \bibinfo{author}{\bibfnamefont{J.}~\bibnamefont{Cox}},
  \bibinfo{journal}{Phys.Lett.} \textbf{\bibinfo{volume}{B517}},
  \bibinfo{pages}{369} (\bibinfo{year}{2001}), \eprint{hep-ph/0006160}.

\bibitem[{\citenamefont{Wetterich}(2014{\natexlab{b}})}]{CWCG}
\bibinfo{author}{\bibfnamefont{C.}~\bibnamefont{Wetterich}}
  (\bibinfo{year}{2014}{\natexlab{b}}), \eprint{1408.0156}.

\end{thebibliography}

\end{document}